\documentclass[journal]{ieee/ieeetj}

\usepackage{cite}
\usepackage{amsmath,amssymb,amsfonts}
\usepackage{algorithmic}
\usepackage{graphicx,color}
\usepackage{textcomp}
\usepackage{xcolor}
\usepackage{hyperref}
\graphicspath{{source/fig/}}
\DeclareGraphicsExtensions{.pdf,.png,.eps}
\usepackage{textcomp}
\usepackage{array,multirow,siunitx,stfloats,svg,wrapfig,colortbl}
\usepackage[caption=false,font=footnotesize]{subfig}
\usepackage[caption=false,font=footnotesize]{subfig}
\ifCLASSOPTIONpeerreview
    \usepackage{censor}
    \usepackage[switch]{lineno}
    \linenumbers

    \AtBeginEnvironment{equation}{\linenomath}\AtEndEnvironment{equation}{\endlinenomath}
    \AtBeginEnvironment{align}{\linenomath}\AtEndEnvironment{align}{\endlinenomath}

\else
    
\fi

\AtBeginDocument{\definecolor{tmlcncolor}{cmyk}{0.93,0.59,0.15,0.02}\definecolor{NavyBlue}{RGB}{0,86,125}}

\def\OJlogo{\vspace{-4pt}$<$Society logo(s) and publication title will appear here.$>$}
\def\seclogo{\vspace{10pt}$<$Society logo(s) and publication title will appear here.$>$}

\def\authorrefmark#1{\ensuremath{^{\textbf{#1}}}}

\title{Double-Profile Intersection (DoPIo) Ultrasound: Pointwise Shear Elasticity Estimation using Paired Confocal Displacement Profiles}
\author{
    Keita~A.~Yokoyama\authorrefmark{1},~Graduate~Student~Member,~IEEE, 
    Md.~Murad~Hossain\authorrefmark{2},~Member,~IEEE, 
    Sabiq~Muhtadi\authorrefmark{1},~Graduate~Student~Member,~IEEE, 
    and~Caterina~M.~Gallippi\authorrefmark{1},~Senior~Member,~IEEE
}
\affil{
    Lampe Joint Department of Biomedical Engineering, University of North Carolina at Chapel Hill and North Carolina State University, Chapel Hill, NC 27599 USA
}
\affil{
    Department of Electrical and Computer Engineering at the University of Hawai'i at M\={a}noa, HI 96822 USA
}
\corresp{
    Corresponding author: Caterina M. Gallippi, email: \mbox{cmgallip@email.unc.edu}
}
\authornote{
    This work was funded by the NCI, NHLBI, and NIDDK of the National Institute of Health (NIH), under award numbers R01HL092944, R01CA281150, R01DK107740, R25DK131344, R43DK112492, and T32DK007750. M. M. Hossain was affiliated with the Lampe Joint Department at the time of his contributions.
}

\begin{document}
\receiveddate{XX Month, XXXX}
\reviseddate{XX Month, XXXX}
\accepteddate{XX Month, XXXX}
\publisheddate{XX Month, XXXX}
\currentdate{XX Month, XXXX}
\doiinfo{XXXX.2025.xxxxxxx}
\markboth{}{Yokoyama \emph{et al.}: DoPIo Ultrasound: Pointwise Shear Elasticity Estimation using Paired Confocal Displacement Profiles}

\begin{abstract}
Current acoustic radiation force (ARF) based methods for quantifying tissue elasticity primarily rely on shear wave propagation. However, their spatial resolution is limited by the need for spatial averaging, and their accuracy is affected by shear wave guidance, out of plane reflections, and geometric dispersion, which reduce their applicability in mechanically complex tissues. This study introduces a novel technique called Double Profile Intersection (DoPIo) ultrasound, which enables pointwise estimation of shear elastic modulus within the region of ARF excitation by leveraging the scatterer shearing rate. This rate is inferred by tracking ARF induced displacement using two tracking beams with different lateral widths. The wider beam captures scatterers located outside the ARF excitation region that begin to displace as shearing propagates. The time at which the two resulting displacement profiles intersect is mapped to shear elastic modulus using an empirically derived model based on finite element simulations. \emph{In silico}, DoPIo estimated shear elastic modulus with a median error of -0.02 kPa and a median absolute deviation of 1.98 kPa in elastic materials up to 35 kPa. Experimental validation \emph{in vitro} and \emph{ex vivo} demonstrated that DoPIo reliably distinguished softer regions from stiffer ones, and its modulus estimates remained consistent across varying ARF push amplitudes, provided sufficient displacement estimation signal to noise ratio. DoPIo offers a feasible approach for high resolution, on axis shear elasticity estimation and holds promise as a quantitative biomarker that is independent of ARF amplitude. 

\end{abstract}
\maketitle
\setcounter{page}{1}

\section{Introduction}
\label{sec:intro}

\IEEEPARstart{T}{issue} elasticity has long served as a biomarker for pathology, from manual palpation to modern approaches in ultrasonic elastography~\cite{nightingale_feasibility_2001}. Elastography techniques estimate mechanical properties of tissue by analyzing how tissue moves in response to internal or external forces. One such external force is acoustic radiation force (ARF), where focused ultrasonic pulses are noninvasively generated~\cite{sigrist_Ultrasound_2017,doherty_Acoustic_2013}. ARF-based elastography has become particularly useful for identifying compositional and structural changes in soft organs including the liver~\cite{frulio_Evaluation_2013, barr_Can_2020}, breast~\cite{magalhaes_Diagnostic_2017, krouskop_Elastic_1998,tsai_Sonographic_2013}, thyroid~\cite{zhang_Acoustic_2015}, and prostate~\cite{krouskop_Elastic_1998}.

Two major categories of ARF based methods are acoustic radiation force impulse (ARFI)~\cite{nightingale_feasibility_2001} and shear wave elasticity imaging (SWEI)~\cite{sarvazyan_Shear_1998}. In ARFI, the amplitude of tissue displacement within the ARF excitation region is inversely related to tissue stiffness. However, displacement amplitude is influenced by both tissue stiffness and the ARF amplitude, which is typically unknown, limiting ARFI to qualitative assessments of relative stiffness within a spatial region assumed to experience consistent ARF amplitude. In SWEI, the velocity of ARF induced shear waves is directly related to tissue stiffness. The velocity must be measured over millimeter scale distances, which requires spatial averaging. Additionally, shear wave propagation is affected by wave guidance, out of plane reflections, and geometric dispersion, all of which complicate interpretation in mechanically complex tissues.

To overcome these limitations, we introduce an alternative acoustic radiation force-based method for tissue elasticity quantification called Double Profile Intersection (DoPIo) ultrasound. Instead of relying on shear wave velocity, DoPIo estimates elastic modulus by analyzing the scatterer shearing rate within the region of ARF excitation.

\section{Double-Profile Intersection Elastography}
\label{sec:theory}

To understand DoPIo imaging, consider that an acoustic radiation force (ARF) excitation induces displacement of sub-resolution scatterers. At the moment of excitation, scatterers located at the center of the ARF point spread function (PSF), where the ARF amplitude is highest, displace more than those near the edges of the PSF, where the amplitude is lower~\cite{sarvazyan_Shear_1998, ostrovsky_NonDissipative_2006,sarvazyan_Acoustic_2021}. Scatterers located outside the ARF PSF do not initially displace. This spatial gradient in displacement creates scatterer shearing, which propagates outward from the center of the ARF PSF over time~\cite{czernuszewicz_Experimental_2013}.

If these scatterer displacements are tracked using a beam with a PSF that is confocal with, and equally wide as the ARF push beam, the resulting A-line will reflect the weighted average displacement of all of the excited scatterers~\cite{walker_fundamental_1995,mcaleavey_Estimates_2003}. On the other hand, if a confocal tracking beam has a PSF that is narrower than the ARF excitation, then the resulting A-line will reflect only the average motion of the scatterers near the center of the ARF PSF. These scatterers respond immediately to the ARF excitation and then undergo elastic recovery. In contrast, if displacements are tracked using a confocal beam with a PSF that is wider than the ARF excitation, the resulting A-line will average together both the initially displaced scatterers and those that begin to move later due to the propagation of scatterer shearing. As a result, the displacement profiles that will be measured from the A-lines generated by these different tracking approaches will differ. 

\begin{figure}[tbp]
    \includesvg[width=\linewidth]{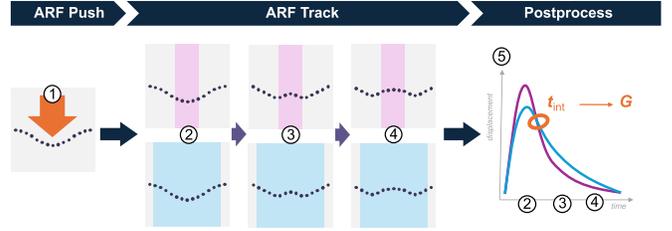}
    \caption{Conceptual illustration of the beam sequence and signal processing employed in DoPIo elastography. From left to right: (1) An acoustic radiation force excitation is applied with a predetermined focal configuration. (2-4) Scatterer displacements are tracked over time using two confocal tracking beams with different lateral widths. (5) The time at which the resulting displacement profiles intersect, denoted as $t_{\textnormal{int}}$, reflects the rate of lateral scatterer shearing and is related to shear elastic modulus through an empirically derived model.}
    \label{fig:mechanism}
\end{figure}

Since all three displacement profiles encode information about the mechanical response of the same region of tissue to the same applied ARF, they can be compared to reveal the speed at which scatterer shearing propagates laterally. For example, because scatterer shearing propagates faster in stiffer media, the time at which the narrow and wide tracking beams report the same average scatterer displacement occurs earlier. To quantify this phenomenon, we define the time at which the two displacement profiles intersect as the “time-intersect”, denoted as $t_{\textnormal{int}}$. This time-intersect is related to shear elastic modulus through an empirically derived model. Figure~\ref{fig:mechanism} illustrates DoPIo's conceptual framework.

We previously introduced DoPIo and demonstrated it for quantifying shear elastic modulus \emph{in silico}~\cite{yokoyama_Doubleprofile_2019,yokoyama_Bessel_2023,muhtadi_Double_2025} and experimentally in select imaging and processing pipelines~\cite{yokoyama_Blind_2020,yokoyama_Assessing_2021,yokoyama_Quantitative_2022}. This work builds on our previous work by identifying the ARF excitation and tracking beam focal configurations and sequences, as well as the empirical model, that yield the most accurate and precise shear elastic modulus estimates. The performance of these optimized DoPIo methods is then evaluated \emph{in silico}, \emph{in vitro}, and \emph{ex vivo}.

\section{Methods}
\label{sec:methods}

\begin{figure}[tb]
    \includegraphics[width=\linewidth]{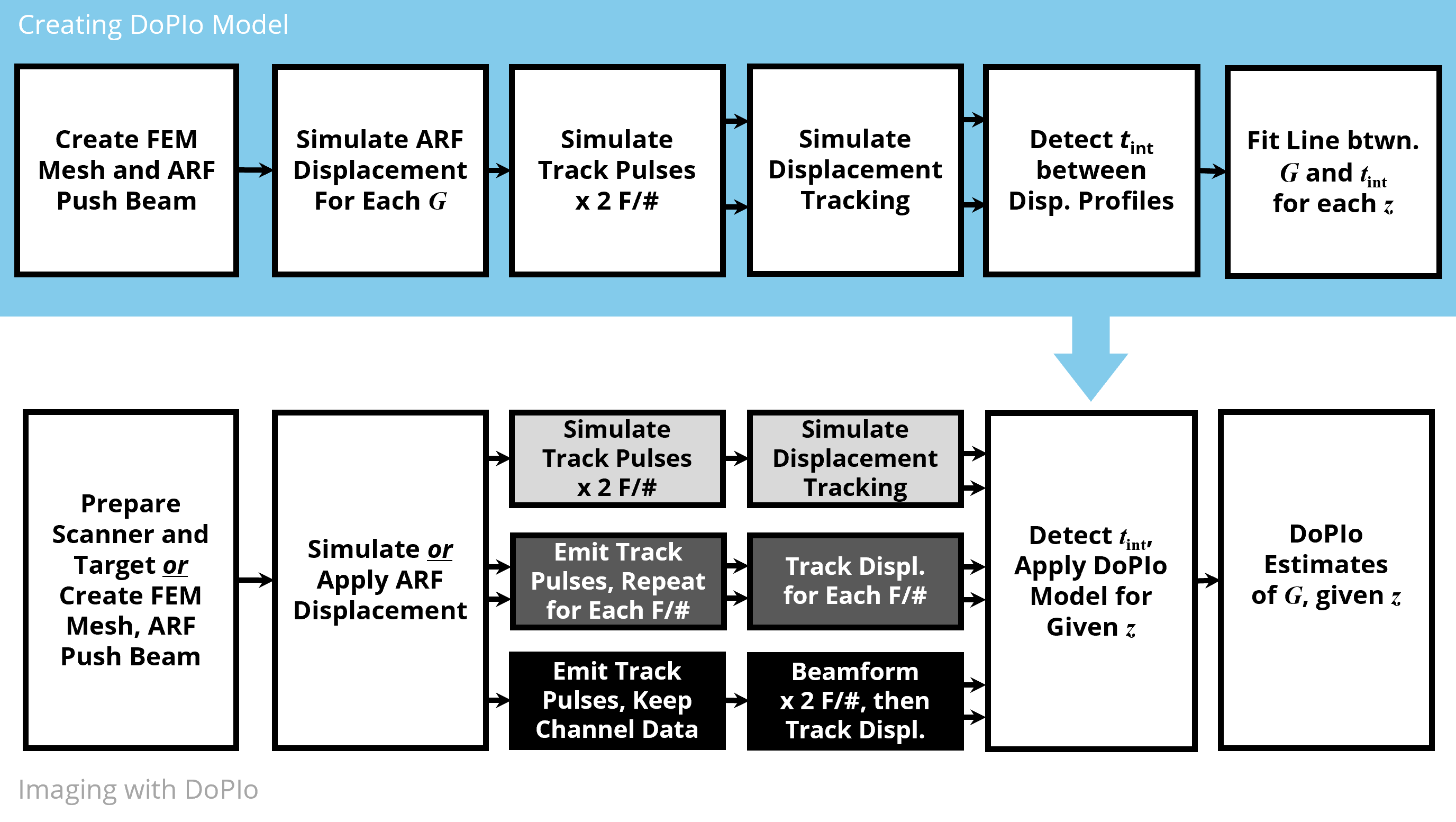}
    \caption{
        Flowchart illustrating the DoPIo imaging procedure, including (top row) development of the empirical model and (bottom row) its evaluation using simulated and experimentally acquired displacement datasets. Times-intersect values ($t_{\mathrm{int}}$) are measured at depth $z$ to estimate the shear elastic modulus ($G$).
        }
    \label{fig:flowchart}
\end{figure}

In the first subsection, the parameters for in silico experiments and the process of detecting $t_{\textnormal{int}}$ are defined. The next subsection describes how empirical models according to Eq.~\ref{eq:modelinvsq} and~\ref{eq:modelarb} are created, as well as how they are evaluated in various additional in silico phantoms. Finally, experimental implementations of DoPIo are demonstrated in a calibrated phantom and in an \emph{ex vivo} porcine liver. The contexts of these procedures are illustrated in Fig.~\ref{fig:flowchart}.

\subsection{Ultrasonic Imaging Simulations}

DoPIo ultrasound was simulated following the procedures originally introduced by Palmeri \emph{et al.}~\cite{palmeri_Experimental_2004,palmeri_Ultrasonic_2006}. Using the linear ultrasound simulator Field-II~\cite{jensen_Calculation_1992,jensen_Field_1997}, ARF excitations were simulated with the specifications of the Philips ATL L7-4 (Bothell, WA, USA), a linear array with an elevational lens focused at 25.0~\si{\milli\meter}. Two ARF focal configurations, F/1.5 and F/3.0, were simulated for comparison, both at a focal depth of 25.0~\si{\milli\meter}. The ARF intensity fields were normalized around the focal depth, scaled to an $I_{\textnormal{SPPA}}$ of 5000~\si{\watt\per\square\centi\meter}, and converted into body forces in the axial direction (i.e. $z$, the direction of acoustic wave propagation) using Eq.~\ref{eq:force}:

\begin{equation}
\vec{F_{z}} = \frac{2 \alpha \vec{I}}{c}
\label{eq:force}
\end{equation}

A sound speed, $c$, of 1540 m/s was assumed, as well as an acoustic attenuation coefficient, $\alpha$, of 0.5 dB/cm/MHz with respect to the transmit frequency. Eq.~\ref{eq:force} was evaluated throughout a hexahedral grid of nodes that were uniformly spaced 0.100 mm apart to represent the ARF push as nodal body forces. This force was applied onto the mesh for 71.2 \si{\micro\second} (300 cycles) at 4.21 MHz, and the resulting displacement response was simulated using the finite element analysis tool LS-DYNA (ANSYS, Canonsburg, PA, USA). Elements were defined as \texttt{*MAT\_ELASTIC} with a density of 1000~\si{\kilo\gram\per\cubic\meter}, Poisson's ratio of 0.499, and Young's moduli based on the experiment described in the following subsections. The outer six elements of the mesh in all directions were replaced with \texttt{*MAT\_PML\_ELASTIC} defined using identical Young's moduli. All non-LS-DYNA operations were performed through MATLAB R2019a (MathWorks Inc., Natick, MA, USA).

Scatterers of normally distributed reflectivity were randomly placed throughout the mesh such that the average scatterer density was 1485~\si{\per\cubic\milli\meter}, corresponding to about 271, 541, and 902 scatterers per resolution cell for F/1.5, F/3.0, and F/5.0 track beams, respectively. Twenty unique realizations of scatterers were simulated. Scatterers were displaced according to the FEM's nodal displacements at every 0.1 ms (i.e. based on a 10 kHz tracking pulse repetition frequency, or PRF), and the displacements were encoded into radio frequency (RF) A-lines using Field-II with two categories of beam sequences. One category involved transmitting a first ARF excitation and tracking it with a narrower tracking beam, then transmitting a second ARF excitation configured as the first and tracking it with a wider tracking beam. We term this sequence a "separate acquisition".  The second category involved transmitting only one ARF excitation and simultaneously tracking it with two tracking beams, one narrower than the other. We term this sequence a "simultaneous acquisition". In the case of the separate acquisition, the tracking beams' focal configuration on transmit matched that on receive. In the case of the simultaneous acquisition, the tracking beam transmit focal configuration matched that of the wider receive focal configuration.

Tracking beam receive focal configurations of F/1.5, F/3.0, and F/5.0 were simulated to support the comparison of DoPIo elastic modulus estimation using the tracking beam pairings (tFc) of F/1.5 and F/3.0 or F/1.5 and F/5.0. These two pairings, each with either separate or simultaneous acquisitions, were combined with either an F/1.5 or F/3.0 ARF excitation, for a total of eight evaluated DoPIo beam sequences (Table~\ref{tbl:fnum}). For all eight beam sequences, tracking beams were two cycles in durations centered at 6.25 MHz. Dynamic receive focusing applied at the center frequency of the track pulses. Using a 10 KHz PRF, ensembles of 6.0 ms were simulated.

\begin{table*}[hbt]
    \centering
    \caption{ARF Push and Track Beam Focal Configuration Combinations (tFc)}
    \begin{tabular}{lllll}
     &
      \multicolumn{2}{c}{Wide Track F/3.0} &
      \multicolumn{2}{c}{Wide Track F/5.0} \\
     &
      \multicolumn{1}{c}{Sequential Acq.} &
      \multicolumn{1}{c}{Simultaneous Acq.} &
      \multicolumn{1}{c}{Sequential Acq.} &
      \multicolumn{1}{c}{Simultaneous Acq.} \\ \hline
    \begin{tabular}[c]{@{}l@{}}F/1.5\\ Push\end{tabular} &
      \begin{tabular}[c]{@{}l@{}}Push F/1.5\\ Track Tx F/1.5, Rx F/1.5 and\\ Track Tx F/3.0, Rx F/3.0\end{tabular} &
      \begin{tabular}[c]{@{}l@{}}Push F/1.5\\ Track Tx F/3.0, Rx F/1.5 and\\ Track Tx F/3.0, Rx F/3.0\end{tabular} &
      \begin{tabular}[c]{@{}l@{}}Push F/1.5\\ Track Tx F/1.5, Rx F/1.5 and\\ Track Tx F/5.0, Rx F/5.0\end{tabular} &
      \begin{tabular}[c]{@{}l@{}}Push F/1.5\\ Track Tx F/5.0, Rx F/1.5 and\\ Track Tx F/5.0, Rx F/5.0\end{tabular} \\ \hline
    \begin{tabular}[c]{@{}l@{}}F/3.0\\ Push\end{tabular} &
      \begin{tabular}[c]{@{}l@{}}Push F/3.0\\ Track Tx F/1.5, Rx F/1.5 and\\ Track Tx F/3.0, Rx F/3.0\end{tabular} &
      \begin{tabular}[c]{@{}l@{}}Push F/3.0\\ Track Tx F/3.0, Rx F/1.5 and\\ Track Tx F/3.0, Rx F/3.0\end{tabular} &
      \begin{tabular}[c]{@{}l@{}}Push F/3.0\\ Track Tx F/1.5, Rx F/1.5 and\\ Track Tx F/5.0, Rx F/5.0\end{tabular} &
      \begin{tabular}[c]{@{}l@{}}Push F/3.0\\ Track Tx F/5.0, Rx F/1.5 and\\ Track Tx F/5.0, Rx F/5.0\end{tabular} \\ \hline
    \end{tabular}
    \label{tbl:fnum}
\end{table*}

To the generated ensembles, one-dimensional axial displacements were tracked using normalized cross-correlation (NCC) with a 2-wavelength (500 \si{\micro\meter}) tracking kernel and a 0.3-wavelength (80 \si{\micro\meter}) search window~\cite{pinton_Rapid_2006}. A polynomial motion filter was applied to each resulting displacement profile based on the pre-ARF push displacement and the last 2.0 ms of displacement estimates~\cite{giannantonio_Comparison_2011} and upsampled tenfold using piecewise-cubic interpolation~\cite{fritsch_Monotone_1980}. 

\subsection{Empirical Model Development}

To generate a library of $t_{\textnormal{int}}$ for materials of different shear moduli, ten LS-DYNA meshes were created. Elements in each mesh were homogeneously assigned one of 10 shear modulus values ranging from 1 kPa in 4 kPa increments up to 37 kPa. For each mesh, ARF-induced displacement profiles were generated as described above, and then $t_{\textnormal{int}}$ were calculated as the first time the paired tracking beams' displacement profiles intersected after the occurrence of their peak displacements. Times-intersect were measured over an axial range that spanned 10 mm above and below the 25 mm focal depth.

At each axial depth, times-intersect underwent one of two options for nonlinear correlation to shear elastic modulus. The first option was an inverse-square transformation, shown in Equation~\ref{eq:modelinvsq}. This transformation represents a treatment of $t_{\textnormal{int}}$ that resembles shear wave-based techniques, where $G$ is proportional to the inverse square of the time duration of shear wave propagation. 

\begin{equation}
G(t_{\textnormal{int}}, z) = A(z) t_{\textnormal{int}}^{-2} + B(z)
\label{eq:modelinvsq}
\end{equation}

The other option, shown in Equation~\ref{eq:modelarb}, generalizes the relationship in Eq.~\ref{eq:modelinvsq} so that an additional degree of freedom is afforded for the exponent of $t_{\textnormal{int}}$, allowing for a more flexible relationship between it and $G$. Here, $\ln\left(x\right)$ is the natural logarithm of $x$, and $t_{\textnormal{ARF}}$ is the duration of the ARF push.

\begin{equation}
\ln\left( G(t_{\textnormal{int}}, z) \right) = A(z) \ln\left( \frac{t_{\textnormal{int}} - t_{\textnormal{ARF}}}{t_{\textnormal{ARF}}} \right) + B(z) 
\label{eq:modelarb}
\end{equation}

For both models, total least squares linear regression~\cite{markovsky_Overview_2007} was performed between transformed times-intersect and shear moduli. This resulted in a map of polynomial coefficients across different axial positions that could be used to convert time-intersect into a shear modulus estimate for a given set of DoPIo beam parameters.

\subsection{DoPIo Performance Evaluation}

To evaluate the performances of the derived empirical models, additional simulations of ARF displacements were performed using FEM simulations of tissues with different stiffnesses for linearly elastic, isotropic materials with shear moduli from 3 to 27 kPa in 4 kPa increments. The beam sequence and model parameters that led to the smallest median error of shear elastic modulus estimates at the focal depth were considered the best-performing process for DoPIo.

Once the best-performing DoPIo beam sequence and model were identified \emph{in silico}, their accuracy and precision for shear elastic modulus prediction were evaluated \emph{in vitro} in a CIRS Type 049 calibrated elasticity phantom (CIRS, Norfolk, VA, USA). Two-dimensional imaging was achieved by sequentially performing DoPIo data collection in 40 lateral positions spaced 0.48 mm apart, leading to a roughly 2 cm lateral field of view. Each of the four cylindrical inclusions in the phantom were imaged in this manner five separate times on a motion isolation table (ThorLabs, Newton, NJ, USA) using a Verasonics Vantage 256 unit (Kirkland, WA, USA) and the L7-4. The ARF voltage was varied from 35 V to 60 V in 5 V increments to evaluate the impact of ARF magnitude on DoPIo modulus estimates. 

DoPIo performance was similarly evaluated ex vivo in a custom porcine liver-based phantom~\cite{hossain_Viscoelastic_2020}. Here, the liver of a pig sacrificed within 15 hours was acquired from a local butcher. A region of tissue away from fascia and large vasculature was identified, and a prism with 3-cm cross-sectional square sides was excised. A 10 mm diameter cylinder was cut using a disposable biopsy punch (Robbins Instruments, Houston, TX, USA). Then, 600 ml of linearly elastic and acoustically attenuating gelatin-graphite material with an approximate Young's modulus of 15 kPa and attenuation of 0.5 dB/cm/MHz was created~\cite{madsen_Tissue_1978,hall_Phantom_1997}. The excised liver was immersed into the solution such that the gelatin solution filled the punched hole and surrounded the liver. The composite phantom was allowed to solidify for 24 hours in a refrigerator. 

The resulting phantom was returned to room temperature, and then three repeated measures were taken using the L7-4 array in an identical manner to that used to image the CIRS phantom. Since the true elasticity of the liver is unknown and the stiffness of the gelatin mixture is only an approximation, a comparison measurement of shear elasticity was taken using the Siemens Acuson Sequoia clinical scanner (Siemens Healthineers Ultrasound Division, Issaquah, WA, USA). An Acuson 10L4 linear array transducer was placed in the identical position as the L7-4 array, and three 2-D shear wave elasticity images were acquired using its standard clinical SWEI imaging mode. These clinical scanner-acquired shear elasticity estimates were used as the validation stanndard for DoPIo modulus estimates.

\section{Results}
\label{sec:results}

\begin{figure}[htbp]
    \centering
    \includegraphics[width=\linewidth]{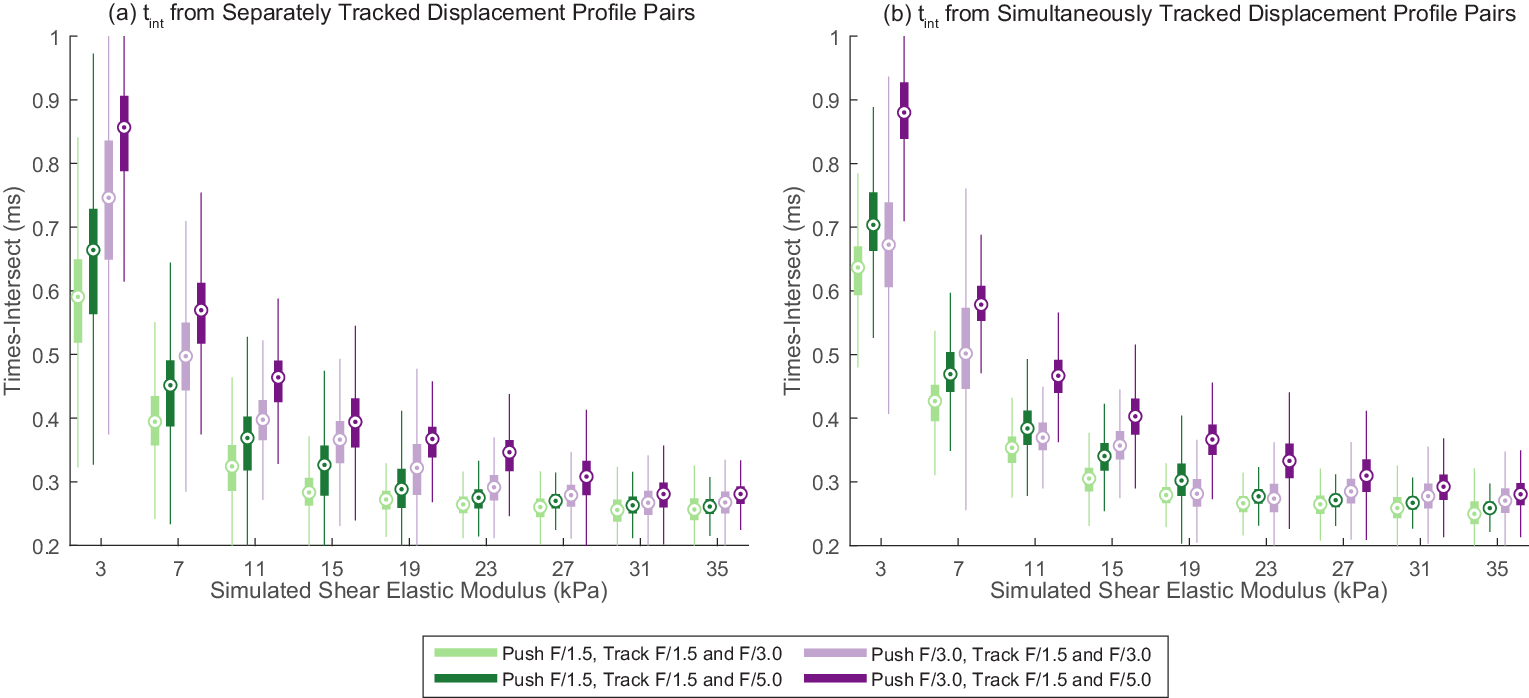}
    \caption{Distributions of times-intersect values for simulated materials with varying elasticities, obtained using (a) separate or (b) simultaneous displacement tracking. Box plots show the median, interquartile range, 95\% confidence interval, and outliers for displacements measured at the focal depth (25 mm) across 20 independent scatterer realizations. Color and shading denote ARF push and tracking beam combinations as indicated in the legend.}
    \label{fig:tintDistrib}
\end{figure}

\begin{figure}[htbp]
    \centering
    \includegraphics[width=\linewidth]{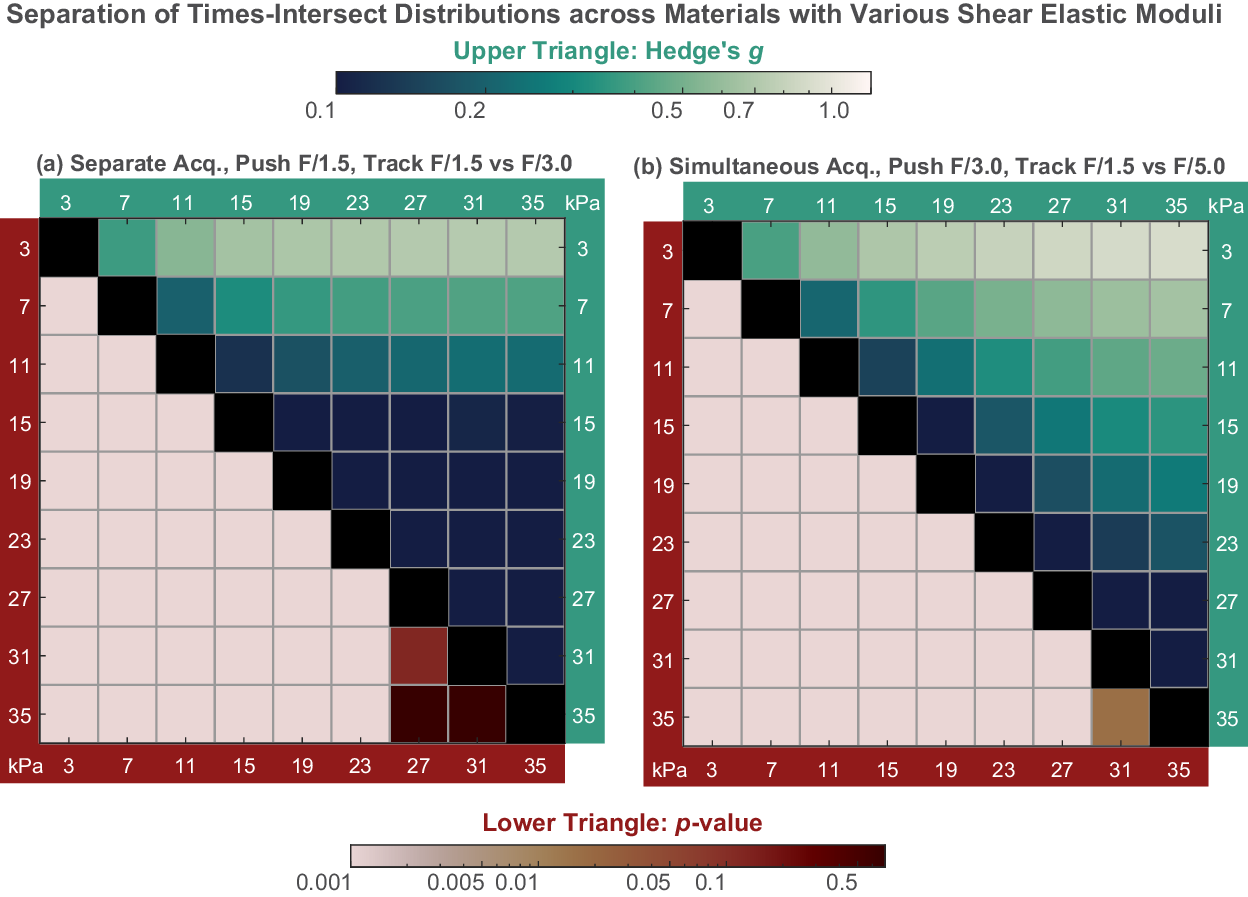}
    \caption{Statistical comparisons of times-intersect distributions between the simulated materials in Fig. 3, with $p$-values (Kruskal-Wallis test with Bonferroni post-hoc corrections) on the lower triangle of each grid and Hedge's $g$ effect sizes on the upper triangle. Results are shown for the focal configuration pairings achieving the greatest separation between materials for (a) separate tracking (F/1.5 push and F/1.5 and F/3.0 track beams) and (b) simultaneous tracking (F/3.0 push and F/1.5 and F/5.0 track beams). }
    \label{fig:tintStat}
\end{figure}

Fig.~\ref{fig:tintDistrib} shows, for the eight evaluated DoPIo beam sequences using the simulated materials that were not included in the empirical model development, distributions of times-intersect measured at a depth of 25.00 mm versus shear elastic modulus. The normalized differences between those distributions (Hedge's $g$ effect sizes) and $p$-values indicating the probability of null hypotheses (Kruskal-Wallis test with Bonferroni posthoc correction) are shown in Fig.~\ref{fig:tintStat}.

\begin{figure}[htbp]
    \includegraphics[width=\linewidth]{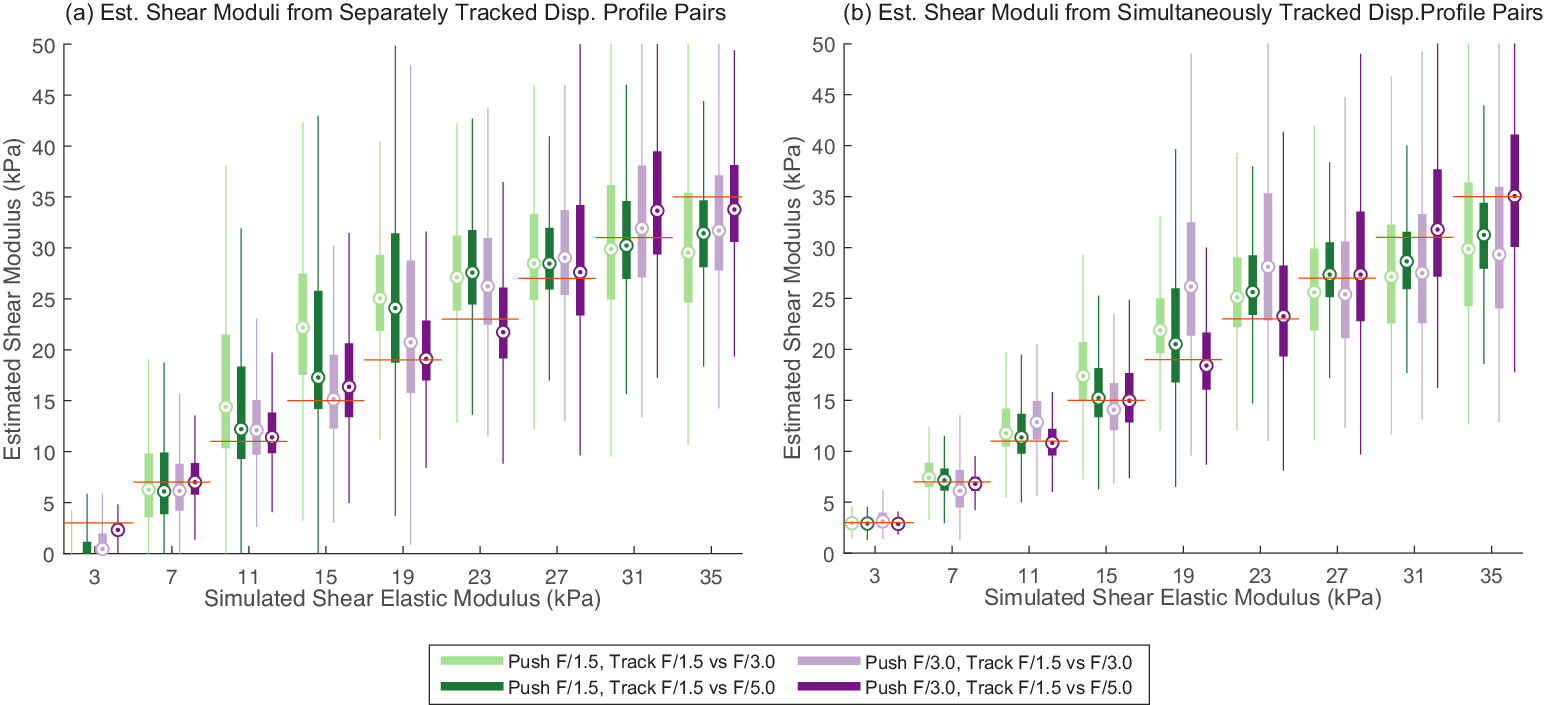}
    \caption{Distributions of DoPIo-derived elastic modulus values for simulated materials with varying elasticities, obtained using (a) separate or (b) simultaneous displacement tracking. Box plots indicate the median, interquartile range, 95\% confidence interval, and outliers for displacements measured at the focal depth (25 mm) across 20 independent scatterer realizations. Color and shading represent ARF push and tracking beam combinations, as indicated in the legend.}
    \label{fig:Gdistrib}
\end{figure}

Table~\ref{tbl:modelr2} lists the coefficients of determination ($R^2$) for the line fits created using each beam sequence and both transformations. For every sequence, the arbitrary power model yielded higher $R^2$ values, so it was selected for modulus estimation \emph{in silico}, \emph{in vitro}, and \emph{ex vivo}. Using this model, Fig.~\ref{fig:Gdistrib} shows shear modulus estimates derived from the times-intersect shown in Fig.~\ref{fig:tintDistrib}. The median and median absolute deviation of shear modulus estimate errors, as well as percent median errors, are shown in Table~\ref{tbl:medianG}. Based on the this table's results, the beam sequence consisting of an F/3.0 ARF excitation with F/1.5 and F/5.0 simultaneous tracking was selected as that which best balanced modulus estimation accuracy and precision and was implemented to demonstrate DoPIo \emph{in vitro} and \emph{ex vivo}.

\begin{table}[htbp]
    \caption{Coefficients of Determination ($R^2$) between True Elastic Modulus and $t_{\mathrm{int}}$ Transforms Assessed at the Focal Depth (25 mm) for Separate versus Simultaneous Tracking across Various Beam Sequences \emph{In Silico}}
    \centering
    \resizebox{0.5\textwidth}{!}{
    \begin{tabular}{lllrr}
    $t_{\mathrm{int}}$ Transform & Push F/\# & tFc & Sep. Track & Simult. Track \\ \hline
    \multirow{4}{*}{\shortstack[l]{Inverse Sq.\\(Eq. 2)}} & \multirow{2}{*}{1.5} & 1.5 and 3.0 & 0.525 & 0.777 \\ \cline{3-5} 
     &  & 1.5 and 5.0 & 0.557  & 0.830 \\ \cline{2-5} 
     & \multirow{2}{*}{3.0} & 1.5 and 3.0 & 0.660 & 0.699 \\ \cline{3-5} 
     &  & 1.5 and 5.0 & 0.836 & 0.803 \\ \hline
    \multirow{4}{*}{\shortstack[l]{Arbitrary Power\\(Eq. 3)}} & \multirow{2}{*}{1.5} & 1.5 and 3.0 & 0.843 & 0.940 \\ \cline{3-5} 
     &  & 1.5 and 5.0 & 0.854 & 0.951 \\ \cline{2-5} 
     & \multirow{2}{*}{3.0} & 1.5 and 3.0 & 0.882 & 0.923 \\ \cline{3-5} 
     &  & 1.5 and 5.0 & 0.943 & 0.954 \\ \hline
    \end{tabular}
    }
    \label{tbl:modelr2}
\end{table}

\begin{table}[htbp]
    \caption{Median \textpm{} Median Absolute Deviation of Errors between True and DoPIo-Derived Shear Elastic Modulus at the Focal Depth (25 mm) for Separate versus Simultaneous Tracking across Various Beam Sequences \emph{In Silico}}
    \centering
    \begin{tabular}{llcc}
    Push F/\#            & tFc        & Sep. Track                                                                  & Simult. Track                                                               \\ \hline
    \multirow{2}{*}{1.5} & 1.5 v. 3.0 & \begin{tabular}[c]{@{}l@{}}0.03 ± 3.63 kPa\\ 0.51 ± 24.19 \%\end{tabular}   & \begin{tabular}[c]{@{}l@{}}0.03 ± 2.39 kPa\\ 0.41 ± 16.51 \%\end{tabular}   \\ \cline{2-4} 
                        & 1.5 v. 5.0 & \begin{tabular}[c]{@{}l@{}}-0.13 ± 3.31 kPa\\ -2.13 ± 20.12 \%\end{tabular} & \begin{tabular}[c]{@{}l@{}}-0.06 ± 2.11 kPa\\ -0.94 ± 13.74 \%\end{tabular} \\ \hline
    \multirow{2}{*}{3.0} & 1.5 v. 3.0 & \begin{tabular}[c]{@{}l@{}}-0.13 ± 3.11 kPa\\ -2.01 ± 20.71 \%\end{tabular} & \begin{tabular}[c]{@{}l@{}}-0.05 ± 3.09 kPa\\ -1.27 ± 19.33 \%\end{tabular} \\ \cline{2-4} 
                        & 1.5 v. 5.0 & \begin{tabular}[c]{@{}l@{}}-0.06 ± 2.53 kPa\\ -1.00 ± 17.03 \%\end{tabular} & \begin{tabular}[c]{@{}l@{}}-0.02 ± 1.98 kPa\\ -0.50 ± 13.02 \%\end{tabular} \\ \hline
    \end{tabular}
    \label{tbl:medianG}
\end{table}

\begin{figure}[htbp]
    \centering
    \includegraphics[width=\linewidth]{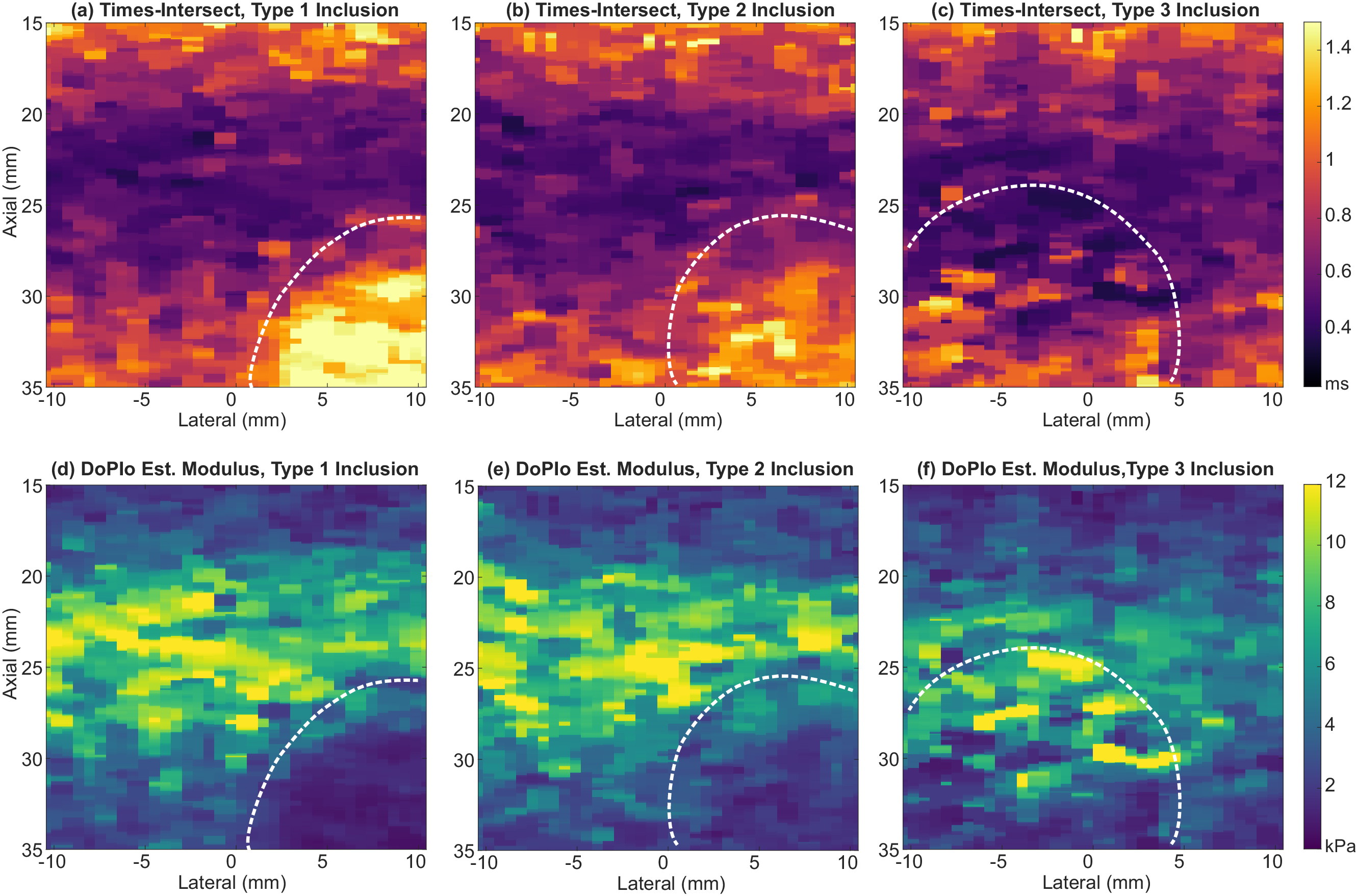}
    \caption{Representative DoPIo images delineating elastic inclusions within a calibrated phantom. Panels show $t_{\mathrm{int}}$ images (a-c) and corresponding DoPIo-derived elastic modulus images (d-f) for inclusions with nominal elasticities of 2.2 kPa (a, d), 5.1 kPa (b, e), and 16.3 kPa (c, f) embedded in an 8.7 kPa background.}
    \label{fig:cirsImage}
\end{figure}

\begin{figure*}[htbp]
    \centering
    \includegraphics[width=\linewidth]{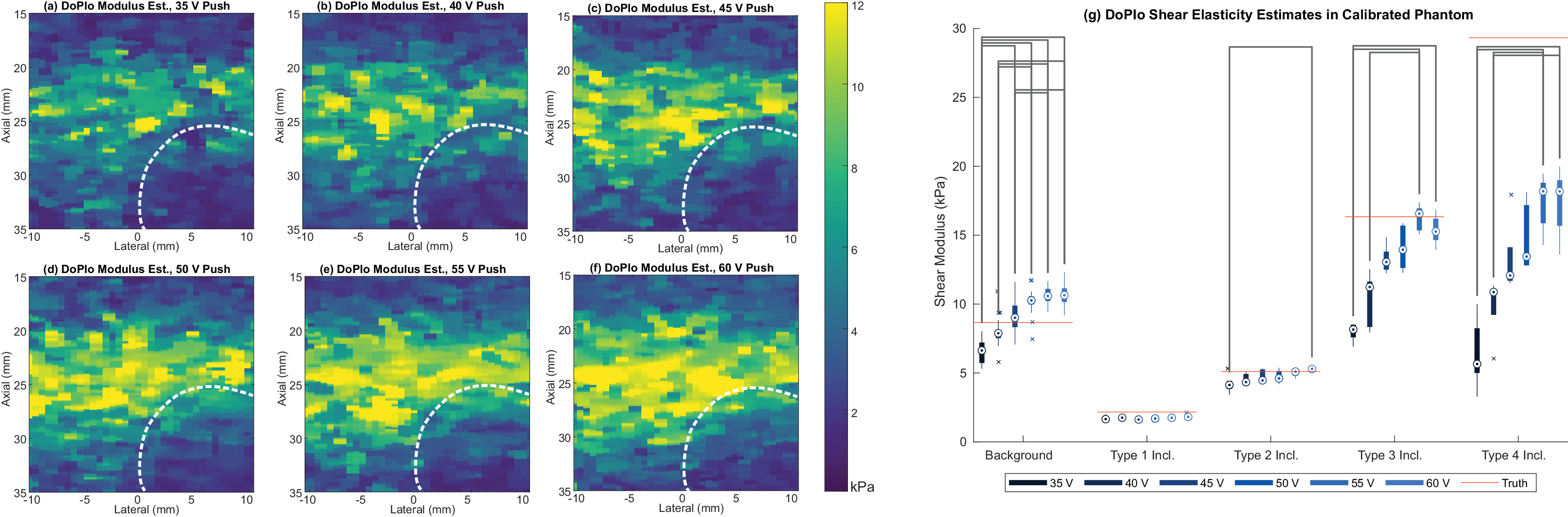}
    \caption{Images of DoPIo-derived elastic modulus for a 5.1 kPa inclusion embedded in an 8.7 kPa background, obtained using ARF push amplitudes ranging from 35 V to 60 V in 5 V increments (a-f). Corresponding distributions of modulus estimates are shown for inclusions with nominal elasticities of 2.2 kPa, 5.1 kPa, and 16.3 kPa within the same 8.7 kPa background. Lines indicate pairs of distributions that differ significantly based on the Kruskal-Wallis test with Bonferroni post hoc correction ($p \leq 0.05$).}
    \label{fig:cirsBoxplot}
\end{figure*}

\begin{figure*}[htbp]
    \centering
    \includegraphics[width=\linewidth]{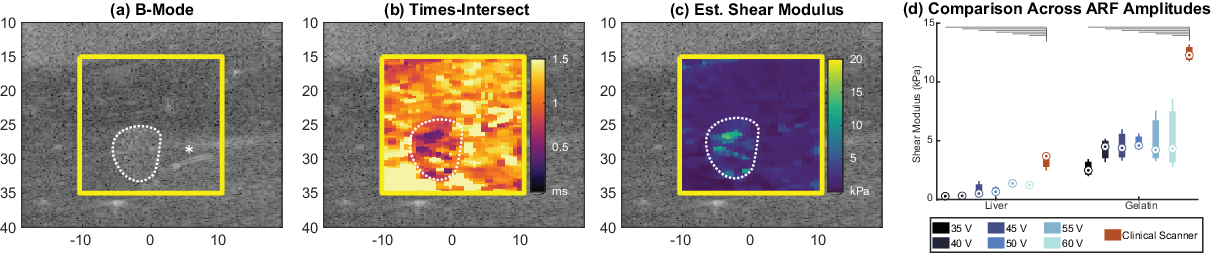}
    \caption{B-mode (a), DoPIo times-intersect (b), and DoPIo-derived elastic modulus (c) images of an ex vivo porcine liver containing a gelatin inclusion. The inclusion boundary, delineated under B-mode guidance, is outlined with white dotted lines. The yellow box in (a) indicates the region used to generate the times-intersect and DoPIo modulus estimate images. Corresponding box plots of DoPIo-derived shear elastic modulus estimates (d). The orange box denotes estimates obtained using the Siemens Sequoia system, and gray lines indicate pairs of distributions that differ significantly based on the Kruskal-Wallis test with Bonferroni post hoc correction ($p \leq 0.05$).}
    \label{fig:liver}
\end{figure*}

Fig.~\ref{fig:cirsImage} depicts, for 55 V ARF excitations, DoPIo images of times-intersect and estimated shear modulus for the background and inclusions in the calibrated elasticity phantom. The estimated median and median absolute deviation of elasticity estimates were 9.0 ± 0.8 kPa in the 8.7 kPa background, 1.6 ± 0.1 kPa in the 2.2 kPa "Type 1" inclusion, 4.5 ± 0.2 kPa in the 5.1 kPa "Type 2" inclusion, and 13.0 ± 0.5 kPa in the 16.3 kPa "Type 3" inclusion, yielding contrast transfer efficiencies of 1.39, 1.19, and 1.30, respectively. The corresponding distributions of estimated moduli for all six employed ARF voltages levels, as well as example images of the Type 2 inclusion across these power levels, are shown in Fig.~\ref{fig:cirsBoxplot}. Fig.~\ref{fig:liver} depicts images of DoPIo times-intersect and estimated shear moduli in the custom liver phantom. The associated distributions of estimated moduli in the background liver tissue and the gelatin inclusion for the six ARF voltage levels are also shown, with the SWEI validation standard indicated in orange. 

\section{Discussion}
\label{sec:discussion}

\subsection{Insights on Times-Intersect and its Measurement}

Fig.~\ref{fig:tintDistrib} shows three important trends. First, across all beam sequences, $t_{\textnormal{int}}$ decreased nonlinearly with increasing shear modulus. For a given change in modulus, differences in $t_{\textnormal{int}}$ were greater for a pair of softer materials versus those of stiffer materials. This trend is important because, since DoPIo estimates modulus from $t_{\textnormal{int}}$, it implies that distinguishing shear elastic moduli in stiffer media is inherently more challenging than among lower ones.

Second, when separate tracking is performed, the dynamic range of $t_{\textnormal{int}}$, as well as differences in their magnitudes between moduli, were greater when the ARF push beam was wider and the difference in tracking beam widths was larger. In particular, differences in $t_{\textnormal{int}}$ between moduli were largest using the beam sequence that used an F/3.0 ARF excitation and F/1.5 and F/5.0 tracking. Relative to the other examined beam sequences, this beam sequence also achieved the largest effect sizes, as shown in Figure~\ref{fig:tintStat}. This trend aligns with the conceptual basis of DoPIo's approach, where shear elastic modulus is estimated by leveraging scatterer shearing. Faster shearing rates are more effectively distinguished when differences in $t_{\textnormal{int}}$ are greater. This can occur when (1) the difference in lateral widths of the two tracking beams is larger, as well as (2) when scatterers displace in distinctly different manners under one beam's PSF versus the other's. This phenomenon was observed when a F/1.5 and F/5.0 track beam combination was used, as those beams had a greater difference in lateral widths compared to the F/1.5 and F/3.0 pair. Paired with a wider F/3.0 ARF push, the F/1.5 track beam primarily captured scatterers near the center of the ARF focus that displaced relatively uniformly at the time of excitation. Meanwhile, the F/5.0 beam applies a heavier weighting to scatterers located beyond the main lobe of the ARF push beam. Those outer scatterers displace later, at a time influenced by the rate of scatterer shearing, leading to a more distinct displacement profile that enhances the measurable difference in $t_{\textnormal{int}}$.

Third, for that sequence involving an F/3.0 push and F/1.5 and F/5.0 tracking, separate and simultaneous tracking yielded comparable $t_{\textnormal{int}}$ distributions (Fig.~\ref{fig:tintDistrib}). Simultaneous tracking offers advantages in DoPIo by doubling the frame rate, halving the number of required ARF excitations, and reducing susceptibility to motion artifacts. However, it requires transmitting a single pulse with a width matching the wider of the two tracking beams, followed by beamforming the resulting channel data to achieve two distinct receive focal configurations. A potential drawback of this approach is the reduced difference in beam widths, which can lead to smaller $t_{\textnormal{int}}$ differences and, consequently, diminished accuracy of modulus estimation - particularly for stiffer materials. This limitation was most evident for the beam sequence using an F/3.0 ARF excitation with F/1.5 and F/3.0 tracking. Since both tracking beams spanned scatterers that were displaced at the time of the ARF excitation and the difference in beam widths was smaller due to the use of simultaneous tracking, stiffness-based trends in $t_{\textnormal{int}}$ were less clear than in the separate tracking case. However, for the beam sequencing using an F/3.0 ARF excitation and F/1.5 and F/5.0 tracking, $t_{\textnormal{int}}$ distributions across elasticities were similar between separate and simultaneous tracking despite the smaller difference in tracking beam widths for the latter case. 

\subsection{Relating Times-Intersect to Shear Elastic Modulus}

The data included in Table~\ref{tbl:modelr2} demonstrate that the arbitrary power law model (Eq.~\ref{eq:modelarb}) consistently yielded higher coefficients of determination compared to the inverse-square model. However, the derived powers varied from 2.1 at the 25-mm focal depth to 1.8 within 10 mm above or below the focus, closely adhering to the inverse-square relationship expected between shear wave propagation time and shear elastic modulus~\cite{sarvazyan_Shear_1998,palmeri_Ultrasonic_2006}. The deviation from an inverse-square relationship could attribute for the effects of the ARF spatial gradient and duration as well as compression waves, mode conversion, and other factors that impact scatterer shear rate before a planar shear wave is fully formed~\cite{landau_Theory_1986,sarvazyan_Shear_1998,ostrovsky_NonDissipative_2006}.

When using the arbitrary power law model, Fig.~\ref{fig:Gdistrib} and Table~\ref{tbl:medianG} show that DoPIo's shear elastic modulus estimates were generally accurate across all beam sequences, with median errors smaller than the simulated modulus step size of 4 kPa. Notably, the sequence combining an F/3.0 ARF excitation with F/1.5 and F/5.0 simultaneous tracking produced the most accurate estimates, consistent with its ability to generate the most distinct $t_{\textnormal{int}}$ distributions. However, modulus estimates from all evaluated beam sequences exhibited lower precision in stiffer materials, as indicated by larger interquartile ranges (IQRs) in the modulus estimates in Fig.~\ref{fig:Gdistrib}. However, this reduced precision is expected since $t_{\textnormal{int}}$ distributions among stiffer materials tend to overlap more than those among softer materials, making them harder to distinguish. While this limitation may be partially addressed by further optimizing $t_{\textnormal{int}}$ detection through improved DoPIo beamforming or increased PRF, an alternative modeling approach may ultimately be required to enhance precision in stiffer materials. Such an approach could include machine learning techniques that directly leverage paired displacement profiles as inputs, rather than relying solely on $t_{\textnormal{int}}$ as an intermediate metric. Initial results using a support vector machine-based regression model have shown promise in improving modulus estimation precision~\cite{rahmouni_Machine_2020}, though further investigation using more advanced techniques is warranted.

\subsection{\emph{In Vivo} and \emph{Ex Vivo} Acquisitions}

Fig.~\ref{fig:cirsImage} shows that elastic inclusions within the calibrated phantom were successfully delineated in the DoPIo images. Qualitatively compared to the background, the 2.2 and 5.1 kPa inclusions exhibited longer $t_{\textnormal{int}}$ and lower DoPIo-estimated elastic moduli while the 16.3 kPa inclusion showed shorter times-intersect and higher estimated moduli. Quantitatively, DoPIo tended to underestimate the modulus of most features, with median and median absolute deviation errors being larger for stiffer materials (e.g. -3.3 ± 0.5 kPa in the 16.3 kPa inclusion) versus softer ones (-0.6 ± 0.2 kPa for the 5.1 kPa inclusion). This discrepancy may stem from increased overlap in $t_{\textnormal{int}}$ distributions in stiffer materials, leading to less accurate and less precise modulus estimates. Another contributing factor could be limitations in the empirical model used to relate $t_{\textnormal{int}}$ to modulus. Since this model was derived \emph{in silico}, it may not be optimally calibrated for \emph{in vitro} applications. This mismatch is potentially due to inconsistencies between simulated and acquired displacement data or greater measurement error in $t_{\textnormal{int}}$ during experimental acquisitions. The latter issue may be particularly pronounced in stiffer materials, which undergo smaller displacements in response to ARF excitation, resulting in displacement magnitudes that approach the inherent variance of the normalized cross-correlation displacement estimator~\cite{mcaleavey_Estimates_2003,dhanaliwala_Assessing_2012}.

Fig.~\ref{fig:cirsBoxplot} investigates the influence of ARF-induced displacement magnitude on DoPIo modulus estimates. Panels (a)-(f) show that the estimated modulus in the 5.1 kPa inclusion remained stable as the ARF voltage increased from 35 V to 60 V. In contrast, the estimated modulus in the 8.7 kPa background rose between 35 V and 50 V, then plateaued from 50 V to 60 V. These trends are echoed in the box plots in panel~(g). DoPIo modulus estimates for the 2.2 kPa inclusion were consistent across all ARF voltage levels. However, for the 16.3 and 29.0 kPa inclusions, modulus estimates increased with ARF voltage from 35 V to 55 V, then stabilized at 55 V and 60 V. These findings suggest that a minimum threshold of ARF-induced displacement is necessary for reliable times-intersect measurements. Furthermore, for a given focal depth and DoPIo beam sequence, stiffer materials require higher ARF push amplitudes to reach this threshold. Importantly, once sufficient displacement is achieved, DoPIo modulus estimates appear to be independent of ARF amplitude, reinforcing DoPIo's potential for quantitative, on-axis, ARF-based elasticity imaging.

Panel (g) also reveals a noteworthy discrepancy: even at the highest ARF push voltage, DoPIo underestimated the modulus of the 29.0 kPa inclusion by approximately 50\%, exceeding the error observed \emph{in silico} in Fig.~\ref{fig:Gdistrib}. This unexpected deviation may result from miscalibration of the empirical $t_{\textnormal{int}}$-to-modulus model for \emph{in vitro} conditions as previously discussed. Alternatively, it may reflect limitations in accurately and precisely measuring low times-intersect in stiff materials during experimental acquisition. To address this, strategies to increase times-intersect, such as enhancing the difference in lateral widths of the paired tracking beams or increasing the PRF, could be explored. However, these adjustments would come with trade-offs: increasing the difference in lateral beam widths would reduce the lateral field of view for 2D DoPIo imaging and degrade radiofrequency signal-to-noise ratio, while the use of higher PRFs would limit imaging depth.

Fig.~\ref{fig:liver} presents DoPIo imaging results from excised liver tissue. As shown in panels (b) and (c), the stiffer gelatin inclusion exhibited lower times-intersect and higher estimated modulus compared to the softer liver background. Panel~(d) displays boxplots of DoPIo modulus estimates across all examined ARF voltage levels for both materials. In general, modulus estimates remained consistent across voltages for both gelatin and liver features, indicating that sufficient ARF-induced displacements were achieved to capture reliable times-intersect measurements. However, modulus estimates were consistently lower than those derived from clinical scanner-based validation standards. This underestimation conflicts with DoPIo's performance in the calibrated phantom, where modulus estimates for media with sub-10 kPa shear moduli were accurate. One possible explanation is the viscoelastic nature of liver tissue. Viscosity may dampen scatterer shearing, leading to longer times-intersect and, consequently, lower estimated moduli. Future work will focus on developing methods to account for, and possibly directly interrogate, viscoelastic effects in DoPIo imaging. To achieve this goal, modified acquisition parameters may be employed to enhance DoPIo's sensitivity to viscous effects, and new data-driven approaches may be explored to exploit viscoelastic behavior that is encoded in displacement profiles.

\section{Conclusion}
\label{sec:conclusion}

This work introduces DoPIo ultrasound as a novel method for estimating shear elastic modulus within the region of ARF excitation. Unlike conventional ARF-based techniques that rely on shear wave propagation, DoPIo leverages the shearing rate of scatterers by quantifying the time when displacement profiles from confocal tracking beams with differing lateral widths intersect. This time-intersect metric is mapped to shear modulus using an empirically derived model.

Simulations demonstrated that DoPIo can accurately estimate shear modulus across a range of elastic materials, with the best performance achieved using an F/3.0 ARF excitation and simultaneous tracking with F/1.5 and F/5.0 beam configurations. \emph{In vitro} and \emph{ex vivo} experiments confirmed DoPIo's ability to delineate elastic inclusions and maintain consistent modulus estimates across varying ARF amplitudes given a sufficient displacement signal-to-noise ratio. Importantly, once a threshold displacement level was reached, DoPIo estimates were independent of ARF amplitude, supporting its potential as a quantitative on-axis, ARF-based, elasticity imaging method.

Despite these strengths, DoPIo exhibited reduced accuracy and precision in stiffer materials, likely due to overlapping time-intersect distributions and limitations in the empirical model derived from simulations. In \emph{ex vivo} liver imaging, DoPIo underestimated modulus relative to clinical standards, potentially due to scatterer shearing dynamics being impacted by the viscoelastic nature of liver tissue.

Future work will focus on refining the empirical model to better account for physical imaging conditions, improving time-intersect measurement accuracy in stiff materials through optimized beam sequencing and increased pulse repetition frequency, and extending DoPIo to viscoelasticity characterization. Additionally, machine learning approaches that leverage full displacement profiles rather than relying solely on time-intersect may enhance precision and generalizability across tissue types. These efforts aim to further establish DoPIo as an ARF amplitude-independent tool for quantitative elasticity imaging in complex tissue environments.


\section*{Acknowledgment}

The authors wholeheartedly thank Siemens Healthineers for technical and equipment support, and the ITS Research Computing division of the University of North Carolina at Chapel Hill for technical support for the \emph{in silico} study.

K.A.Y., M.M.H., and C.M.G. are inventors on a patent in the United States of America related to this work (9,043,156) and have potential future financial interests in the DoPIo technology. C.M.G. is a current member of the Scientific Advisory Board (SAB) of Verasonics, Inc. but receives no royalties, gifts, or in-kind support in association with her service therein.

This work was funded by the NCI, NHLBI, and NIDDK of the National Institute of Health (NIH), under award numbers R01HL092944, R01CA281150, R01DK107740, R25DK131344, R43DK112492, and T32DK007750. The content is solely the responsibility of the authors and does not necessarily represent the official views of the National Institutes of Health.

\bibliographystyle{IEEEtran}
\bibliography{references}

\end{document}